\documentclass{phostproc}

\title{Photometric detection of internal gravity waves in early-type stars observed by CoRoT}
\author{D. M. Bowman,$^{1}$
          C.~Aerts,$^{1,2}$
          C.~Johnston,$^{1}$
          M.~G.~Pedersen,$^{1}$
          T.~M.~Rogers,$^{3,4}$
          P.~V.~F.~Edelmann,$^{3}$
          S.~Sim{\' o}n-D{\' i}az,$^{5,6}$
          T.~Van~Reeth,$^{7,8}$    
          B.~Buysschaert,$^{1,9}$
          A.~Tkachenko,$^{1}$
          S.~A.~Triana$^{10}$ }

\affiliation{$^{1}$ Instituut voor Sterrenkunde, KU Leuven, Celestijnenlaan 200D, 3001 Leuven, Belgium \\
$^{2}$ Department of Astrophysics, IMAPP, Radboud University Nijmegen, NL-6500 GL Nijmegen, The Netherlands \\
$^{3}$ School of Mathematics, Statistics and Physics, Newcastle University, Newcastle-upon-Tyne NE1 7RU, UK \\
$^{4}$ Planetary Science Institute, Tucson, AZ 85721, USA \\
$^{5}$ Instituto de Astrof{\' i}sicade Canarias, E-38200 La Laguna, Tenerife, Spain \\
$^{6}$ Departamento de Astrof{\' i}sica, Universidad de La Laguna, E-38205 La Laguna, Tenerife, Spain \\
$^{7}$ Sydney Institute for Astronomy (SIfA), School of Physics, The University of Sydney, NSW 2006, Australia \\
$^{8}$ Stellar Astrophysics Centre, Department of Physics and Astronomy, Aarhus University, Ny Munkegade 120, DK-8000 Aarhus C, Denmark \\
$^{9}$ LESIA, Observatoire de Paris, PSL Research University, CNRS, Sorbonne Universit{\' e}s, UPMC Univ. Paris 06, Univ. Paris Diderot, Sorbonne Paris Cit{\' e}, 5 place Jules Janssen, F-92195 Meudon, France \\
$^{10}$ Royal Observatory of Belgium, Ringlaan 3, B-1180 Brussels, Belgium }
            
\shorttitle{Internal gravity waves in massive stars}
\shortauthors{D. M. Bowman \textit{et al.}}  % to be used for contributions with 3 authors and more.

\abs{Early-type stars are predicted to excite an entire spectrum of internal gravity waves (IGWs) at the interface of their convective cores and radiative envelopes. Numerical simulations of IGWs predict stochastic low-frequency variability in photometric observations, yet the detection of IGWs in early-type stars has been limited by a dearth of high-quality photometric time series. We present observational evidence of stochastic low-frequency variability in the CoRoT photometry of a sample of O, B, A and F stars. The presence of this stochastic low-frequency variability in stars across the upper main-sequence cannot be universally explained as granulation or stellar winds, but its morphology is found to be consistent with predictions from IGW simulations.}

\begin{document}

\maketitle

% % % % % % % % % % % % % % % % % % % % % % % % % % % % % % % % % % % % % % % % 

\section{Introduction}
\label{section: intro}

Massive stars play a significant role in stellar and galactic evolution theory, yet their interior rotation, mixing and angular momentum transport mechanisms are poorly constrained \citep{Maeder_rotation_BOOK, Meynet_rotation_BOOK, Aerts2018c*}. Only detailed observational constraints of stellar interiors provided by asteroseismology offer the ability to mitigate uncertainties in current stellar models \citep{ASTERO_BOOK, Aerts2018c*}. Until now, well-characterised interior radial rotation profiles have been found to be quasi uniform in main sequence stars with spectral types between late-B and early-F (see e.g., \citealt{Kurtz2014, Saio2015b, VanReeth2016a, VanReeth2018a, Ouazzani2017a, Aerts2017b}). Clearly a strong angular momentum transport mechanism must be at work within intermediate-mass stars to explain their near-rigid radial rotation rates. 

A promising candidate for the missing angular momentum transport mechanism is internal gravity waves (IGWs), which are travelling waves that are stochastically excited at the interface of a convective region and a stably stratified zone. It has been shown numerically and theoretically that IGWs are efficient at transporting angular momentum and chemical mixing within stars of various masses and evolutionary stages (see, e.g. \citealt{Press1981a, Talon1997b, Pantillon2007, Charbonnel2005a, Lecoanet2013, Rogers2013b, Rogers2015, Rogers2017c}). These IGWs are predicted to produce stochastic low-frequency variability in velocity and temperature spectra near the surface of a star \citep{Rogers2013b, Rogers2015, Edelmann2018a**}. 

However, there have only been a few inferred detections of IGWs in observations (e.g. \citealt{Aerts2015c, Aerts2017a, Aerts2018a, Simon-Diaz2018a, Bowman2018c*}) because of the requirement of high-precision, long duration and continuous photometry from space telescopes. More observational studies of early-type stars are required to address the large shortcomings in the theory of angular momentum transport that have been discovered when comparing observations of main sequence and evolved stars \citep{Tayar2013, Cantiello2014, Eggenberger2017a, Aerts2018c*}. 

Here we discuss the results from \citet{Bowman2018c*} who performed a search for IGWs in CoRoT photometry of O, B, A and F stars, and provided a direct comparison to predictions from numerical simulations of IGWs \citep{Rogers2013b, Rogers2015, Edelmann2018a**}. Furthermore, we demonstrate the negligible systematic differences for an ensemble of early-type stars when characterising stochastic low-frequency variability using amplitude spectra compared to power density spectra. Specifically, we demonstrate that the choice of using amplitude or power density spectra does not resolve the discrepancy between the observed morphology of the stochastic low-frequency variability in early-type stars with the morphology predicted by the granulation scaling law derived from solar-like oscillators.

% % % % % % % % % % % % % % % % % % % % % % % % % % % % % % % % % % % % % % % % 

\section{Characterising stochastic low-frequency variability in photometry}
\label{section: results}

Low-frequency stochastic variability with periods of order days is seemingly ubiquitous in massive stars \citep{Balona1992c, Walker2005a, Rauw2008, Blomme2011b, Buysschaert2015}, especially when considering the low Fourier noise levels obtained from continuous, long-term space photometry. This low-frequency variability, sometimes referred to as red noise, has been demonstrated to be of an astrophysical origin and can arise from different non-mutually exclusive physical phenomena, which are can either be dominant or negligible dependent on a star's location in the HR~diagram. These phenomena include: 
\begin{itemize}
\item granulation (e.g. \citealt{Michel2008a, Kallinger2010c, Chaplin2013c, Hekker2017a});
\item stellar winds (e.g. \citealt{Puls2006, Rami2018a, Krticka2018d});
\item damped pulsation modes (e.g. \citealt{Dziembowski2008, Aerts2011, Aerts2015c}).
\end{itemize}

%	%	%	%	%	%	%	%	%	%	%	%	%	%	%

	\subsection{Morphology of photometric red noise}
	\label{subsection: red noise}
	
	For a systematic search for IGWs across the Hertzsprung--Russell (HR) diagram, \citet{Bowman2018c*} assembled a sample of 35 stars with spectral types between O and early-F observed by CoRoT \citep{Auvergne2009}. Each star has a light curve which ranges in length between $\sim$25 and $\sim$115~d with a median cadence of approximately 32~s \citep{Michel2006c}. For stars in the sample with $S/N \geq 4$ coherent pulsation modes \citep{Breger1993b}, iterative pre-whitening was performed to extract peaks using a non-linear least-squares fit to the light curve (see, e.g. \citealt{Degroote2009a, Kurtz2015b, Bowman_BOOK}) with the cosinusoid model of:
	\begin{equation}
	\Delta m = A \cos \left( 2\pi\nu\left(t - t_0 \right) + \phi \right) ~ ,	
	\label{equation: cosine}
	\end{equation}
	\noindent where $A$ is the amplitude (mmag), $\nu$ is the frequency (d$^{-1}$), $t$ is the time (d), $t_0$ is the zero-point of the time scale and taken to the middle of the light curve, and $\phi$ is the phase (rad). After removing the significant peaks via iterative pre-whitening, \citet{Bowman2018c*} used the residual light curves to calculate power density spectra in which the morphology of the low-frequency stochastic variability was fitted using the Lorentzian function:
	\begin{equation}
	\alpha \left( \nu \right) = \frac{ \alpha_{0} } { 1 + \left( \frac{\nu}{\nu_{\rm char}} \right)^{\gamma}} + P_{\rm w} ~ ,	
	\label{equation: red noise}
	\end{equation}
	\noindent where $\alpha_{0}$ is a scaling factor and represents the amplitude at zero frequency, $\gamma$ is the gradient of the linear part of the model in a log-log plot, $\nu_{\rm char}$ is the characteristic frequency and $P_{\rm w}$ is a frequency-independent (i.e. white) noise term. The characteristic frequency (i.e. the frequency at which the amplitude of background equals half of $\alpha_0$) is the inverse of the characteristic timescale: $\nu_{\rm char} = (2\pi\tau)^{-1}$. 
		
	\begin{figure*}
	\centering
	\includegraphics[width=0.99\columnwidth]{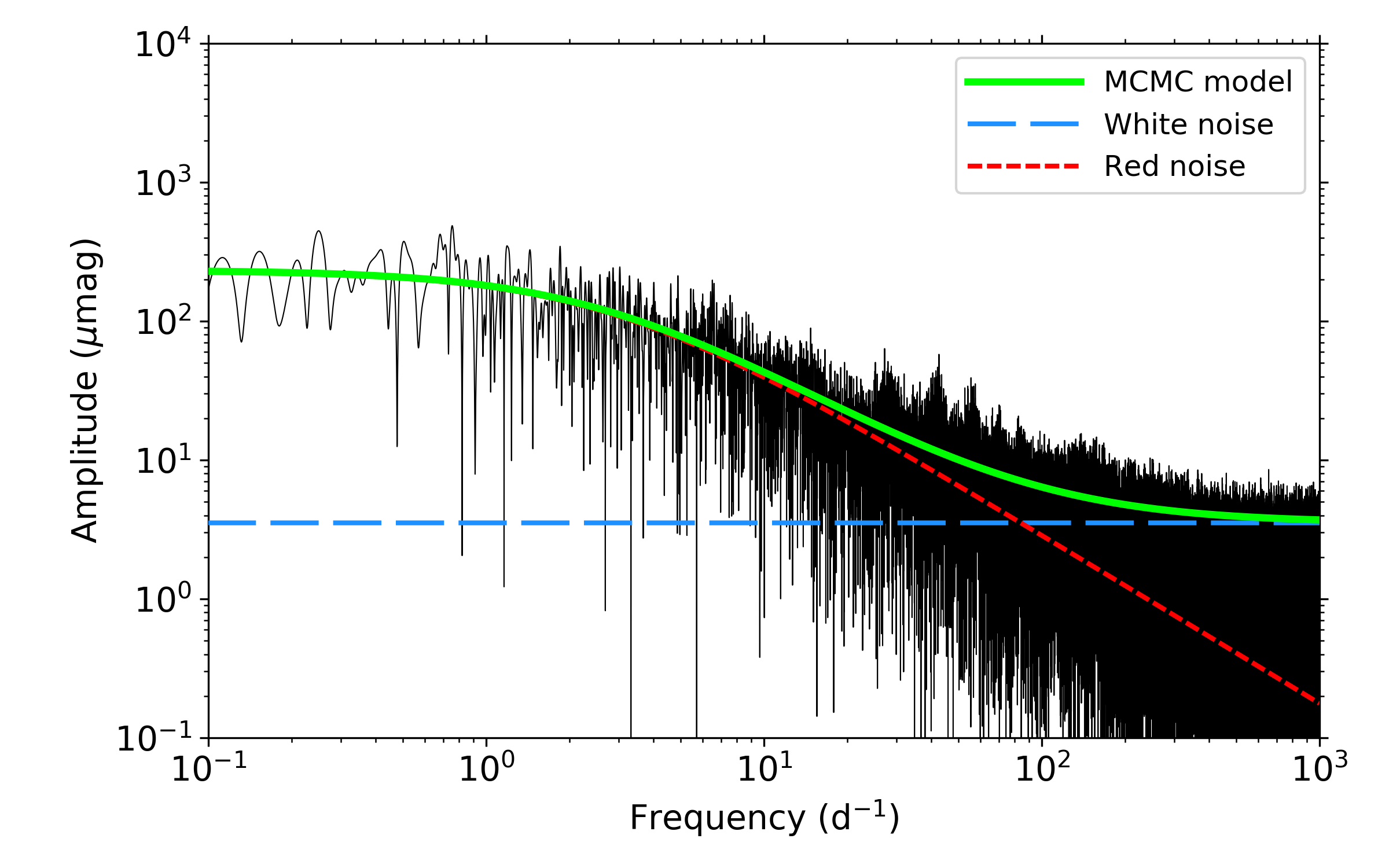}
	\includegraphics[width=0.99\columnwidth]{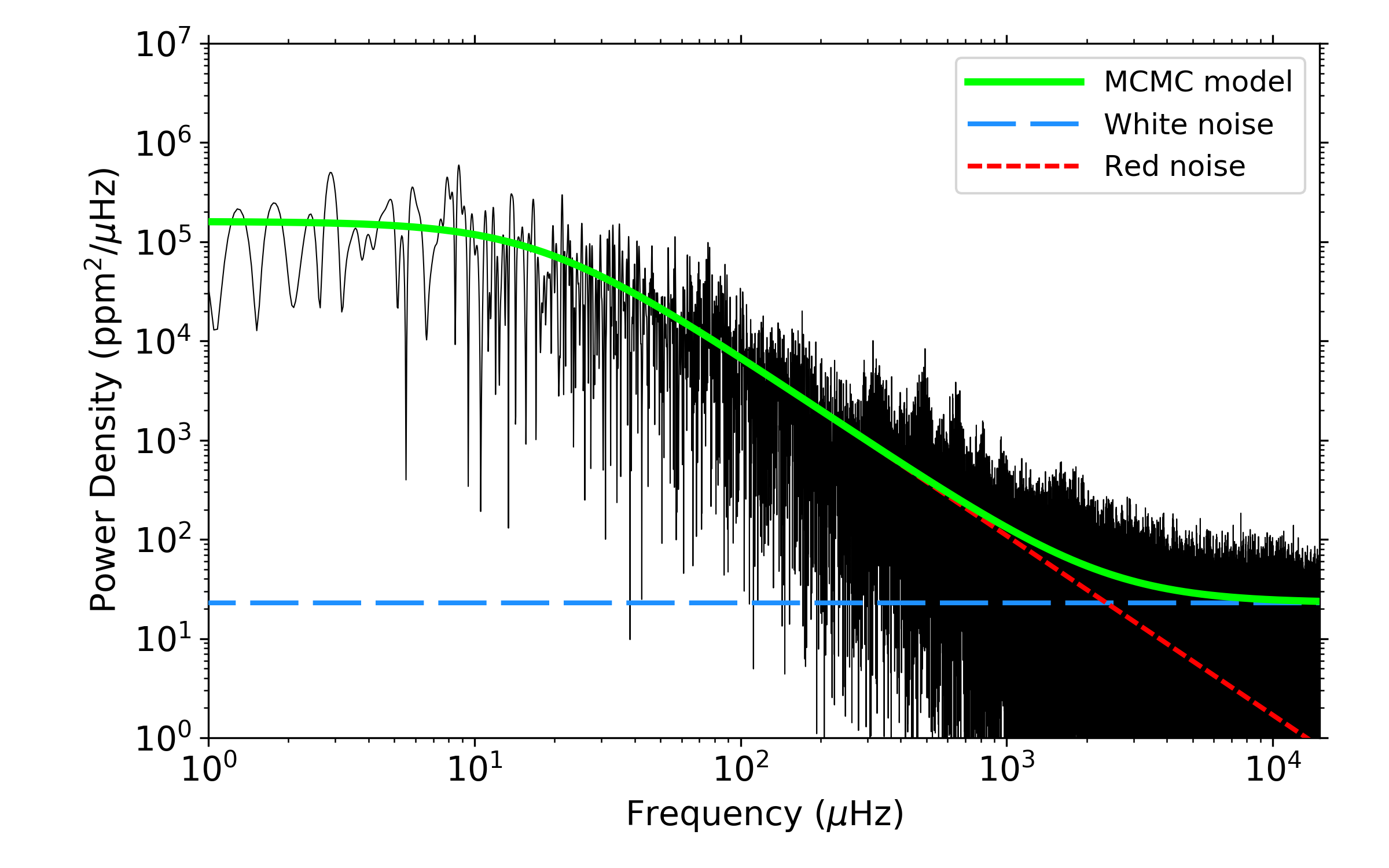}
	\includegraphics[width=0.99\columnwidth]{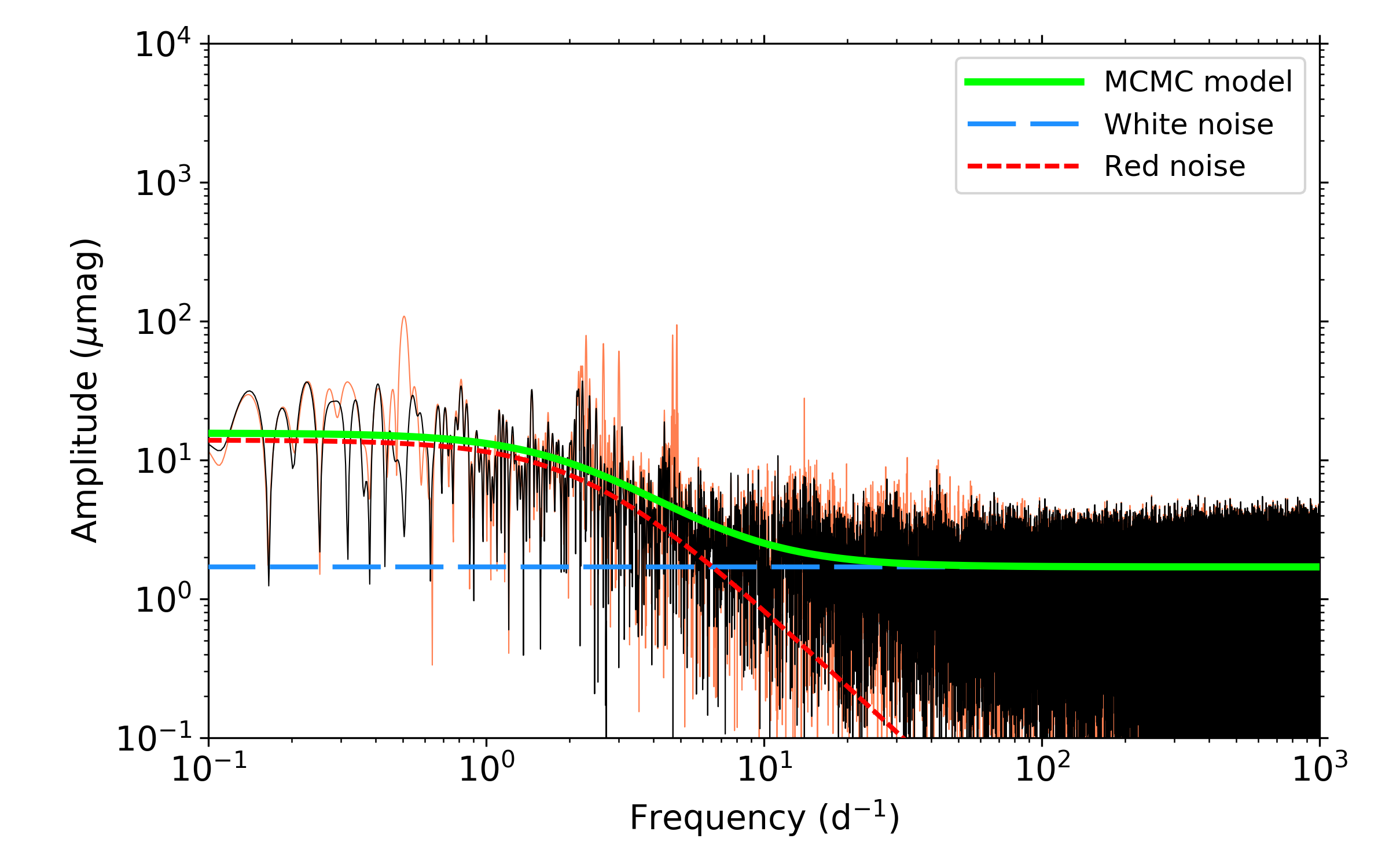}
	\includegraphics[width=0.99\columnwidth]{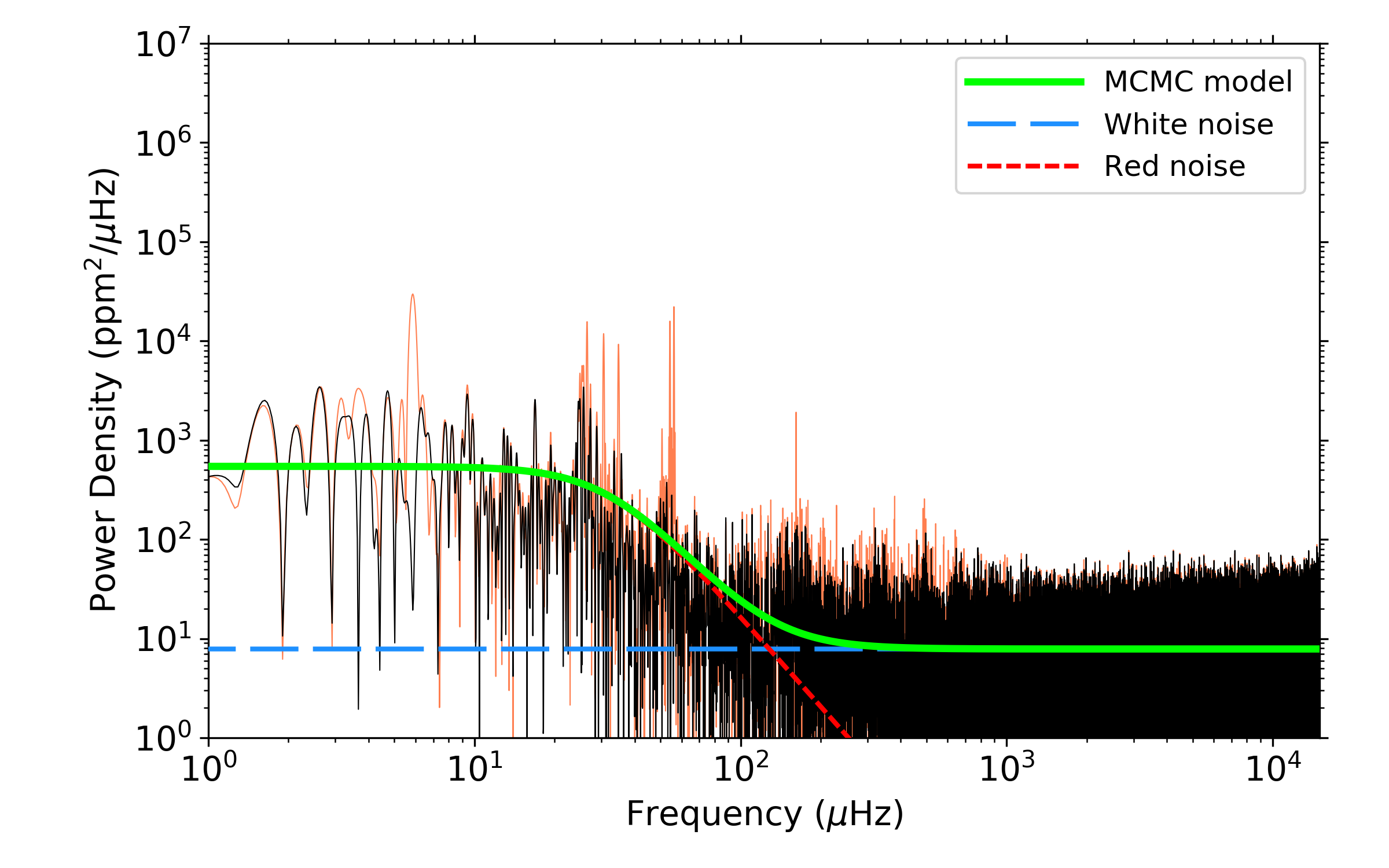}
	\includegraphics[width=0.99\columnwidth]{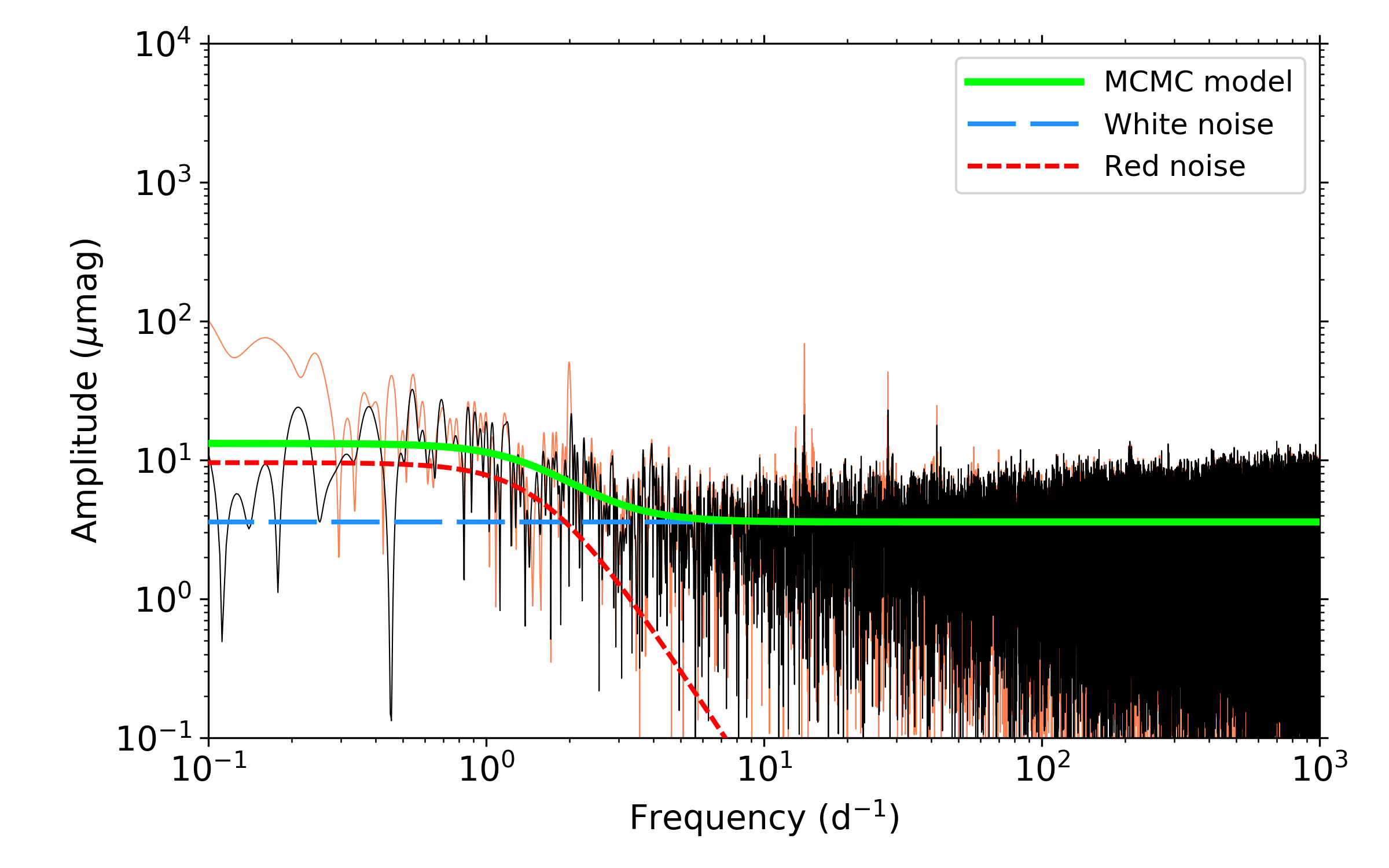}
	\includegraphics[width=0.99\columnwidth]{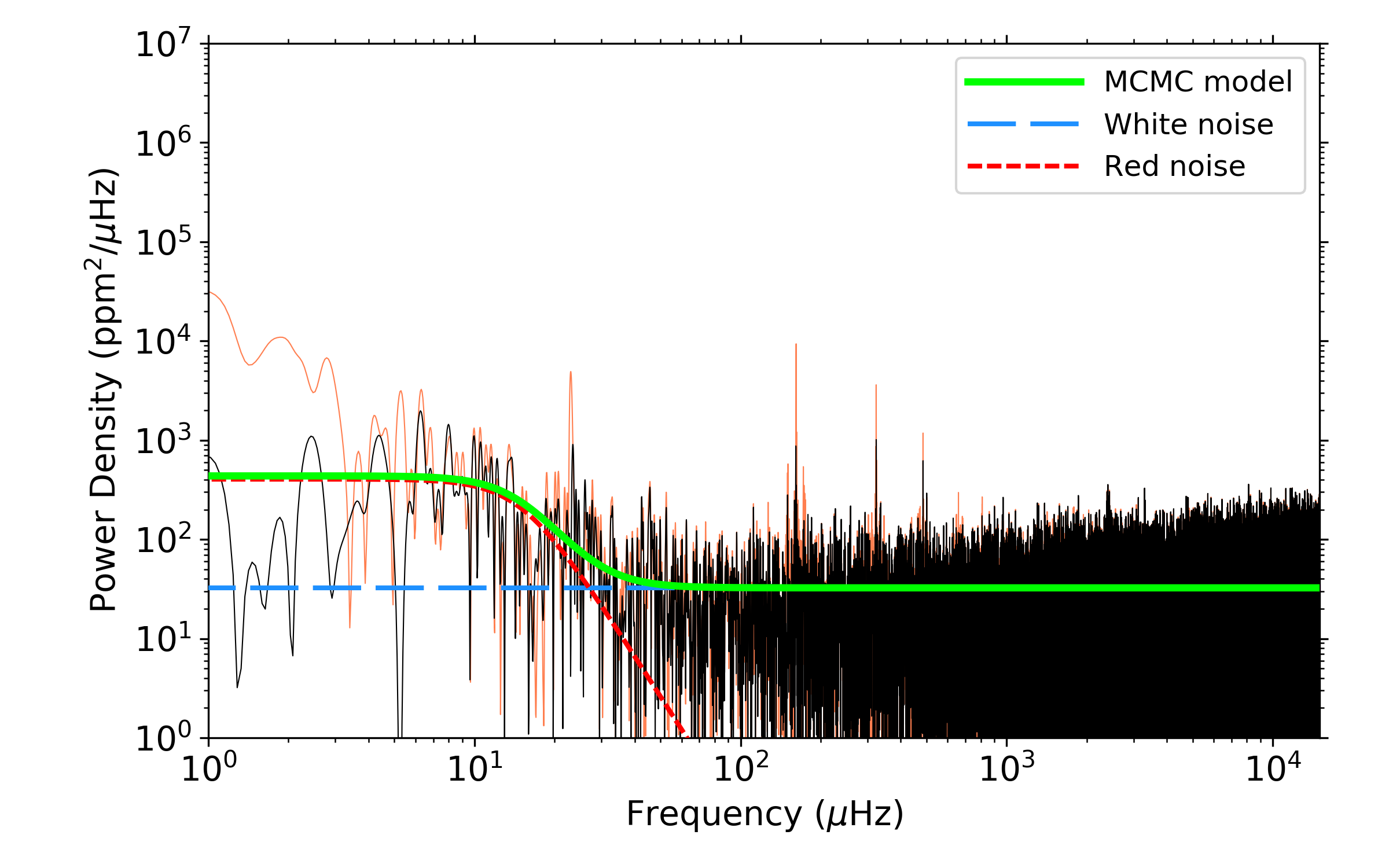}
	\includegraphics[width=0.99\columnwidth]{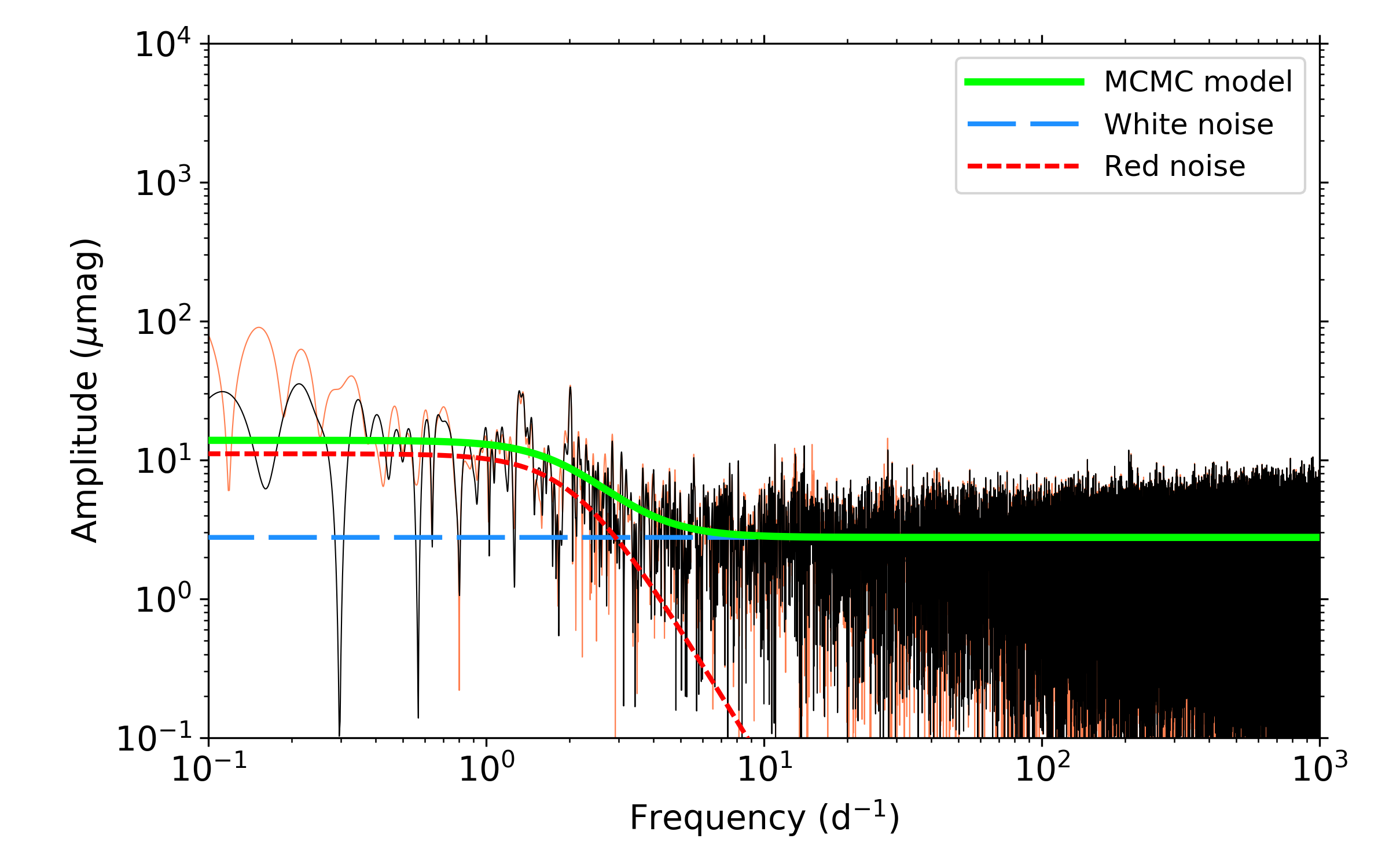}
	\includegraphics[width=0.99\columnwidth]{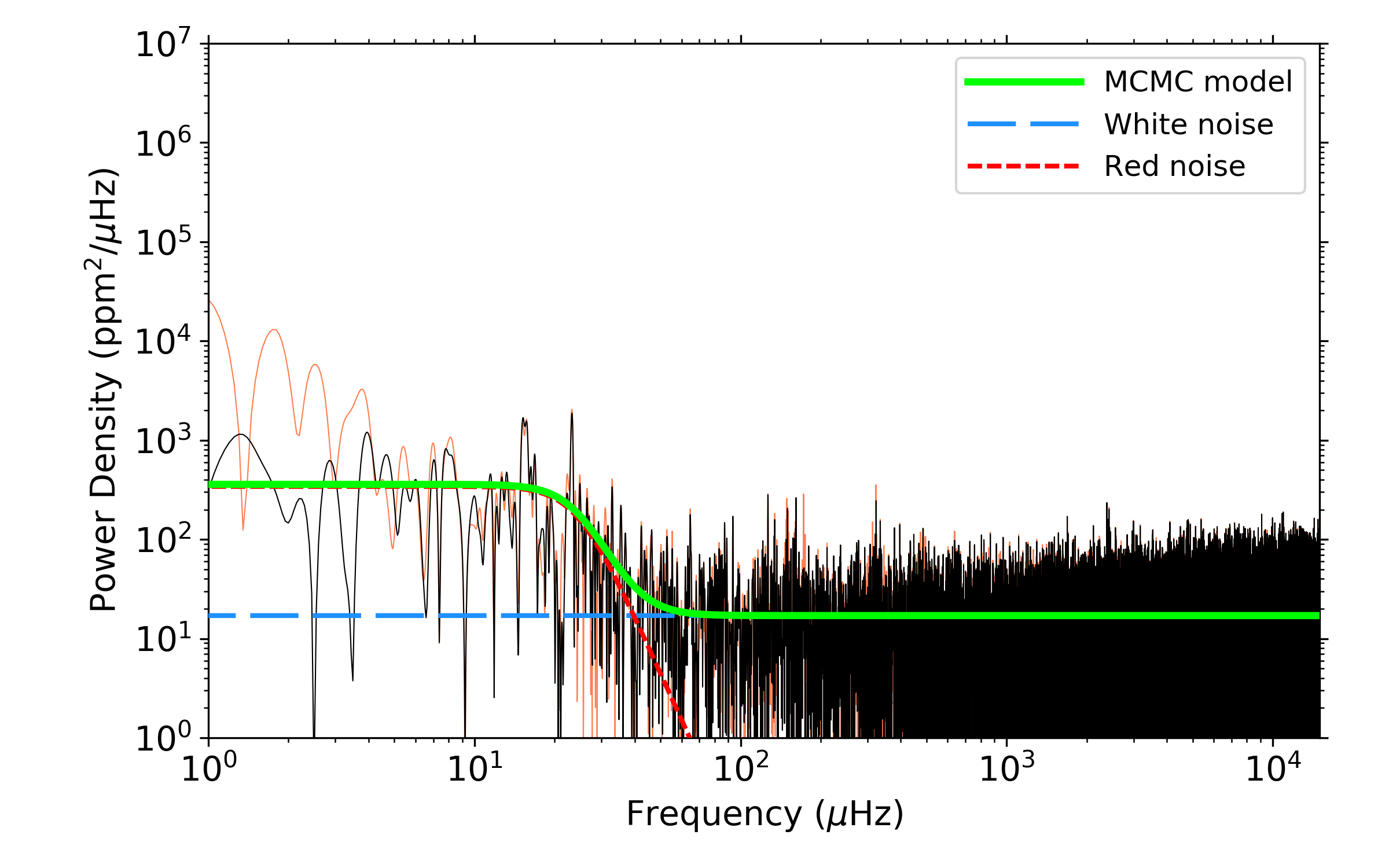}
	\caption{From top to bottom: the logarithmic spectra of the O4\,V star HD~46223, the O9.2\,V star HD~46202, the B9.5\,V star HD~174967, and the A3\,V star HD~174990 using CoRoT observations. The left and right columns show amplitude and power density spectra, respectively. For stars in which pre-whitening was applied (e.g. HD~174967, HD~46202, and HD~174990), the spectrum of the original light curve is shown in orange, and the spectrum of the residual light curve is shown in black. Since no significant peaks exist in the spectra of HD~46223, only the spectra of the original light curve are shown in black. The best fit in each panel was obtained using Eq.~(\ref{equation: red noise}) to each spectrum is shown as the solid green line. This figure is adapted from \citet{Bowman2018c*}.}
	\label{figure: examples}
	\end{figure*}
	
	We show examples of the resultant best-fit models using Eq.~(\ref{equation: red noise}) of four stars, the O4\,V star HD~46223, the O9.2\,V star HD~46202, the B9.5\,V star HD~174967, and the A3\,V star HD~174990, using amplitude spectra and power density spectra in the left and right columns of Fig.~\ref{figure: examples}, respectively. For stars which underwent pre-whitening of significant (i.e. $S/N \geq 4$) frequencies from their original light curves, the spectra of the original and residual light curves are shown in orange and black, respectively. Clearly, the CoRoT photometry of main-sequence stars with spectral types that range between mid-O to early-A reveals significant low-frequency stochastic variability that is well-characterised by a Lorentzian in an amplitude or power density spectrum, which given the large parameter space of the stars cannot be unanimously explained by granulation or stellar winds.

	%	%	%	%	%	%	%	%	%	%	%	%	%	%	%

	\subsection{Excluding granulation for early-type stars}
	\label{subsection: granulation}
	
	Originally discussed in detail by \citet{Kjeldsen1995}, the characteristic granulation frequency can be estimated using the scaling relation:
	\begin{equation}
	\nu_{\rm gran} \propto \frac{c_s}{H_p} \propto M~R^{-2}~T_{\rm eff}^{-1/2} ~ ,
	\label{equation: granulation}
	\end{equation}
	\noindent where $\nu_{\rm gran}$ is the granulation frequency, $c_s$ is the sound speed, $H_p$ is the pressure scale height, $M$ is mass, $R$ is radius, and $T_{\rm eff}$ is the effective temperature. As demonstrated by \citet{Kallinger2014} for thousands of solar-like oscillators, a tight correlation exists between $\nu_{\rm gran}$ and the stellar parameters, such that Eq.~(\ref{equation: granulation}) is valid for stars on the red giant branch. Similarly, from a Lorentzian fit to the power density spectra of the evolved A stars HD~50844 and HD~174936, \citet{Kallinger2010c} demonstrated that the characteristic frequencies, $\nu_{\rm char}$, of the red noise in these stars were also consistent with granulation scaled to higher masses. However, two evolved A stars are hardly representative of all early-type stars, with no similar pulsating A stars, i.e. $\delta$~Sct stars, observed in the {\it Kepler} data set (see e.g., \citealt{Balona2014b, Bowman2018a}). If the use of the granulation scaling relation in Eq.~(\ref{equation: granulation}) is physically justified for the (sub-)surface convective zones of intermediate- and high-mass stars, one would expect a similar smooth and tight correlation between the measured $\nu_{\rm char}$ and the stellar parameters.
	
	\begin{figure*}
	\centering
	\includegraphics[width=0.49\textwidth]{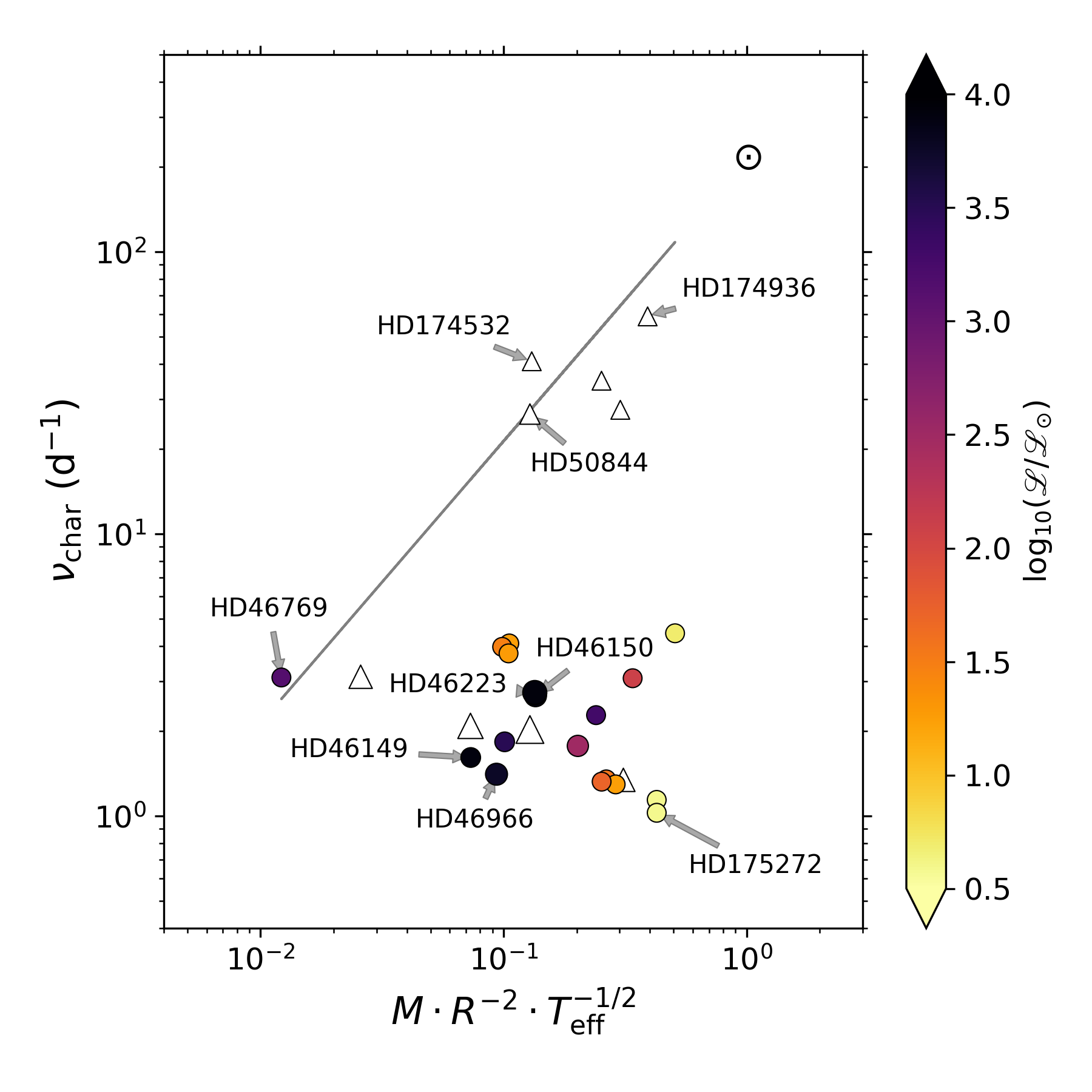}
	\includegraphics[width=0.49\textwidth]{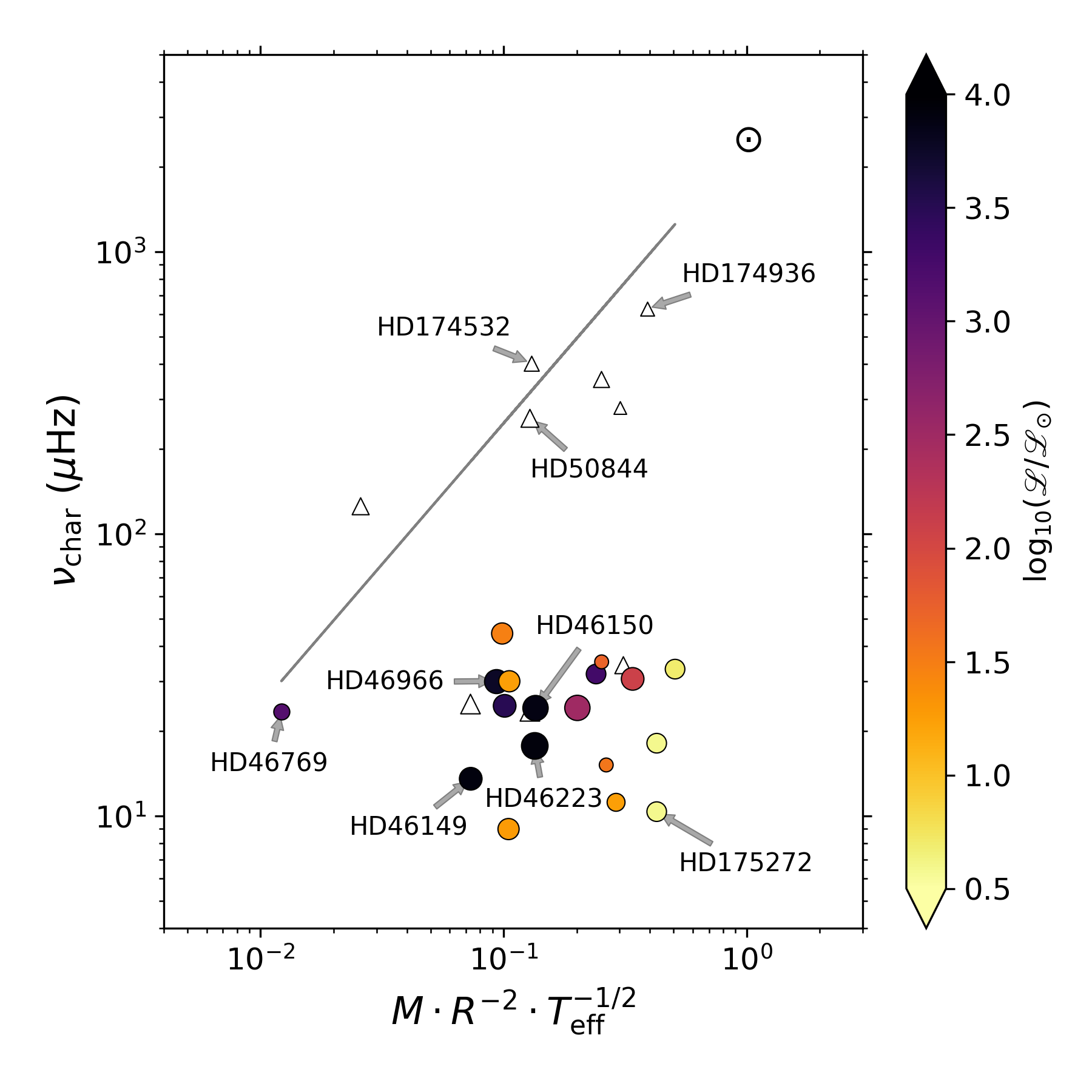}
	\caption{Morphology of stochastic low-frequency variability using amplitude spectra (left panel) and power density spectra (right panel). The distribution of measured characteristic frequencies, $\nu_{\rm char}$, as a function of stellar parameters are given in solar units. The grey line represents the granulation frequency scaled using Eq.~(\ref{equation: granulation}) with the Sun's location indicated by the $\odot$ symbol. The circles have been colour-coded by a star's spectroscopic luminosity and the size of the symbol is proportional to the fitting parameter, $\alpha_0$. Stars with insufficient CoRoT data for a robust determination of $\gamma$ are shown as white triangles. This figure is adapted from \citet{Bowman2018c*}.}
	\label{figure: granulation}
	\end{figure*}
	
	To investigate if the observed low-frequency variability observed in their sample of early-type stars could be explained by granulation, \citet{Bowman2018c*} found that the majority of stars had measured $\nu_{\rm char}$ values that were an order of magnitude smaller than the predicted characteristic granulation frequency as predicted by Eq.~(\ref{equation: granulation}) for their spectroscopic parameters. In Fig.~\ref{figure: granulation}, we summarise the variance in the morphology of the observed red noise across the sample of early-type stars, and specifically how the morphology of the stochastic low-frequency variability in these stars depends on the stellar parameters. In Fig.~\ref{figure: granulation}, the distribution of measured $\nu_{\rm char}$ values as a function of the stellar parameters is shown, with the solid grey line indicating the predicted granulation frequency of the Sun scaled to different stars, for which the left and and right panels correspond to if the best-fitting model using Eq.~(\ref{equation: red noise}) was applied to amplitude spectra or power density spectra, respectively. Stars with insufficient time spans of CoRoT data to resolve the complex beating patterns of close-frequency pulsation modes (see e.g. \citealt{Bowman2016a}), or complex window patterns that bias the best-fitting model are shown as white triangles in Fig.~\ref{figure: granulation}. We refer the reader to \citet{Bowman2018c*} for a thorough discussion of the sample of stars being plotted in Fig.~\ref{figure: granulation}.
	
	For stochastic low-frequency variability, it is typical to use Fourier spectra in ordinate units of power density (e.g. ppm$^2$/$\mu$Hz) as opposed to amplitude (e.g. ppm or mmag), such that limitations and biases imposed by different stars having light curves of different lengths are avoided. Although the differences in the best-fit parameters from Eq.~(\ref{equation: red noise}) can be significantly different depending on whether the choice is made to use amplitude spectra or power density spectra for an individual star, it is clear that the ordinate units of the spectrum has little impact on the systematic disagreement between the measured $\nu_{\rm char}$ values and the expected granulation frequencies for the ensemble of early-type stars shown in Fig.~\ref{figure: granulation}. Therefore, irrespective or using amplitude spectra or power density spectra, the majority of early-type stars in Fig.~\ref{figure: granulation} have $\nu_{\rm char}$ values that are approximately an order of magnitude smaller than expected for granulation, thus their stochastic low-frequency variability appears to be inconsistent with granulation.

	%	%	%	%	%	%	%	%	%	%	%	%	%	%	%

	\subsection{Photometric evidence of IGWs}
	\label{subsection: IGWs}

\begin{figure}
	\centering
	\includegraphics[width=0.49\textwidth]{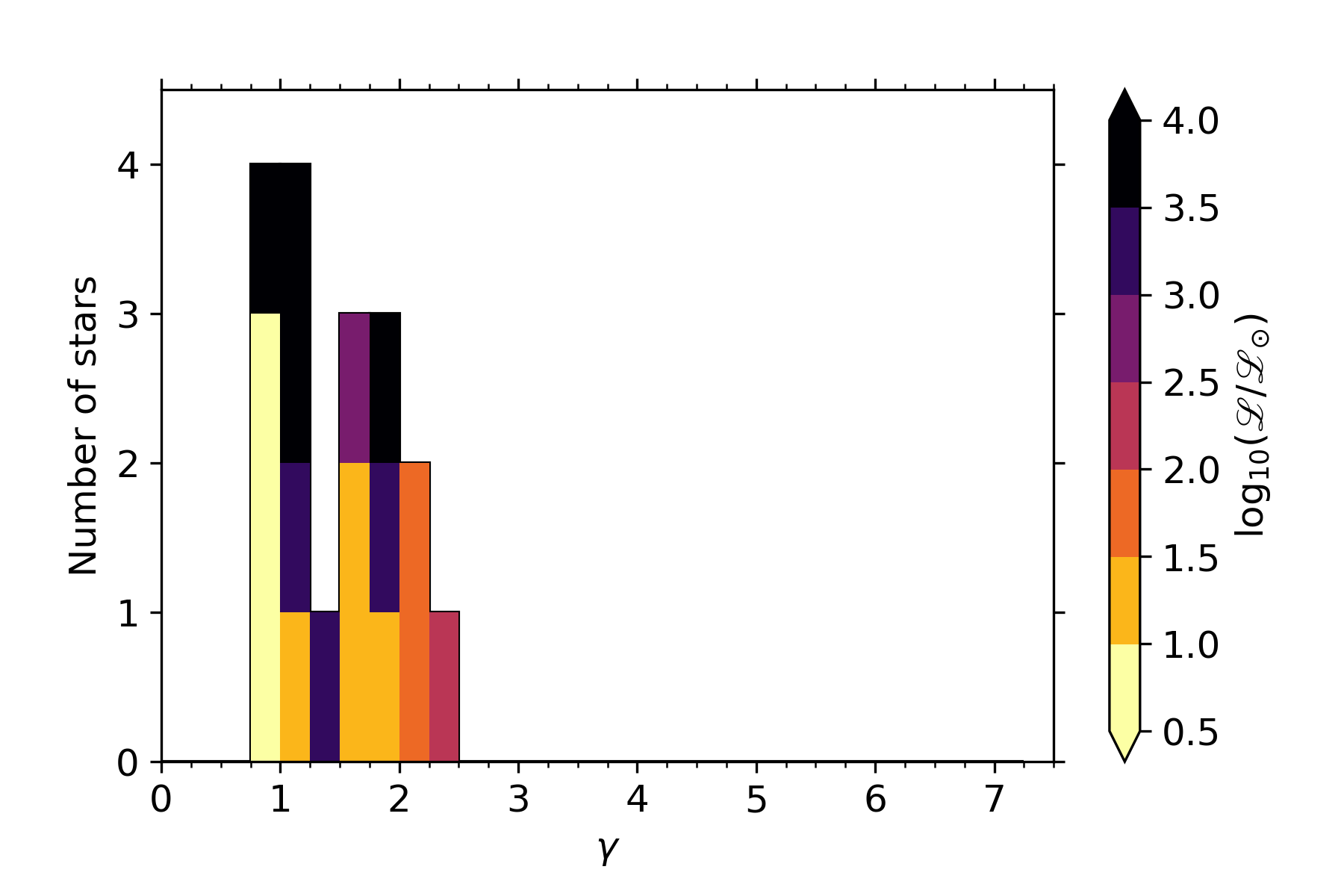}
	\includegraphics[width=0.49\textwidth]{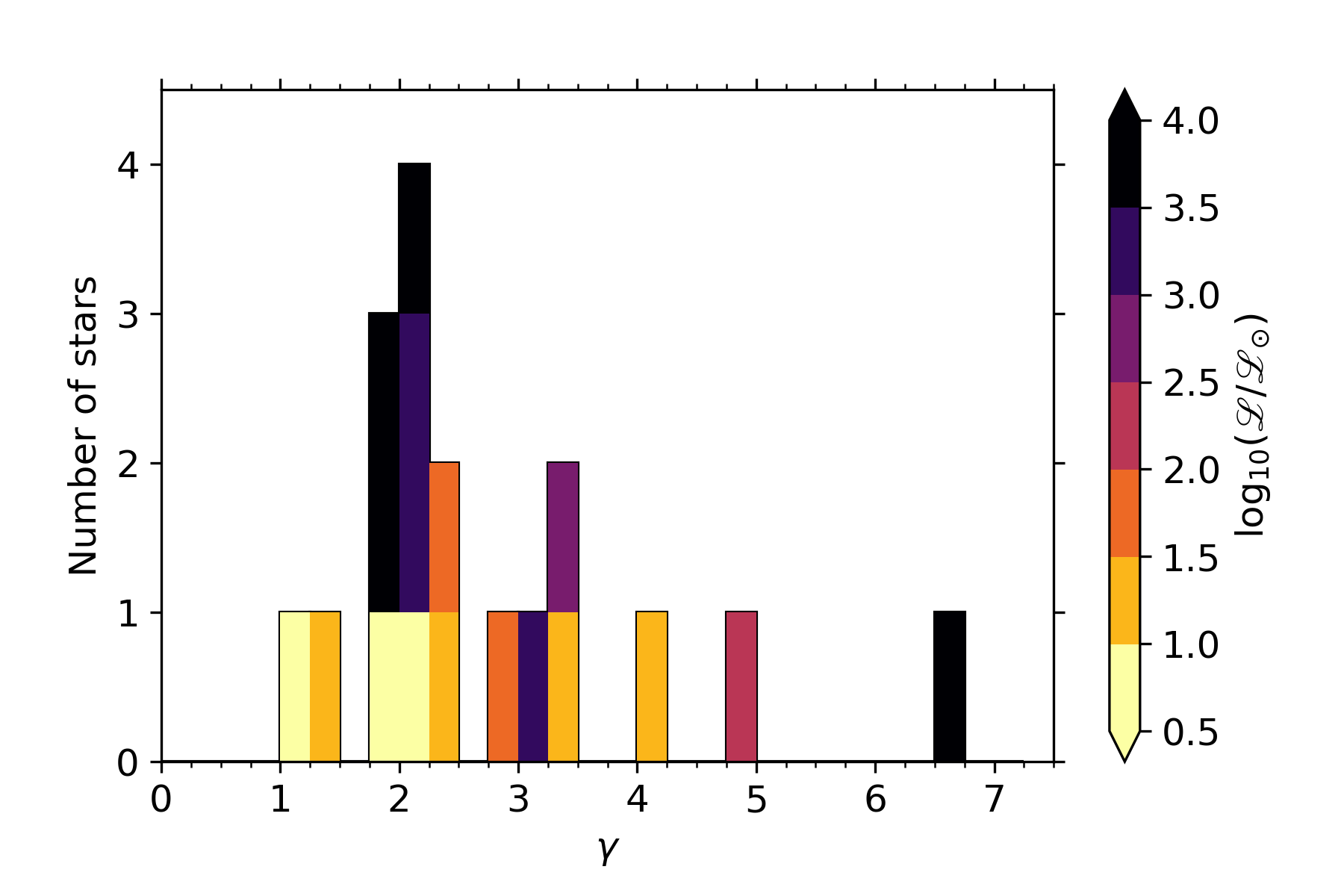}
	\caption{The distribution of $\gamma$ values (c.f. Eq.~\ref{equation: red noise}) for amplitude (top panel) and power density spectra (bottom panel).}
	\label{figure: IGW}
	\end{figure}

	Neither strong stellar winds nor granulation are able to ubiquitously explain the presence of stochastic low-frequency variability across spectral types ranging from mid-O to early-A. On the other hand, IGWs are predicted to be excited in all stars on the main sequence with a convective core, i.e. $M \gtrsim 1.5$~M$_{\odot}$. Numerical simulations of IGWs predict the spectrum of the low-frequency variability to have an exponent approximately between $-1$ and $-3$ depending on the rotation rate and interior differential rotation profile of a star \citep{Rogers2013b, Rogers2015, Edelmann2018a**}. The distribution of the measured $\gamma$ values (c.f. Eq.~\ref{equation: red noise}) when applied to amplitude and power density spectra are shown in the top and bottom panels of Fig.~\ref{figure: IGW}, respectively. Almost all stars in the sample have low-frequency stochastic variability characterised by a Lorentzian with $\gamma \leq 5$. 
	
	As demonstrated in Fig.~\ref{figure: granulation}, the stochastic low-frequency variability observed in early-type stars appears inconsistent with granulation. This is also supported by the difference in the scatter of $\gamma$ values for amplitude and power density spectra shown in Fig.~\ref{figure: IGW}. The disparity in measured $\nu_{\rm char}$ values from the granulation scaling relation shown in Fig.~\ref{figure: granulation} when using power density spectra and the conclusion that the observed stochastic low-frequency variability is not caused granulation is further supported by the narrow range in $\gamma$ values in Fig.~\ref{figure: IGW} obtained using amplitude spectra. The small range in observed $\gamma$ values implies a common underlying physical mechanism that is driving the observed low-frequency variability, which is linked to temperature variability at the surface of the star and observed in photometric amplitude spectra.
		
	Recent theoretical work and 3D numerical simulations of IGWs triggered by turbulent convection in a Cartesian model by \citet{Couston2018b} predict a similar power-law frequency spectrum as that of \citet{Lecoanet2013}, with a frequency exponent of $-13/2$ (corresponding to $\gamma = 6.5$). This is much steeper than the power-law predicted by 3D simulations of core convection in a 3-M$_{\odot}$ star by \citet{Edelmann2018a**}, which take into account the spherical configuration, as well as an appropriate density stratification from a realistic stellar model. The 3D spherical simulations of \citet{Edelmann2018a**} predict an IGW frequency spectrum with a frequency exponent between $-1.5$ and $-3$ (i.e. $\gamma \in [1.5, 3]$), which is similar to the simulations of core convection by \citet{Rogers2013b} and \citet{Augustson2016a}. The analysis of CoRoT stars with convective cores leads to $\gamma \leq 5$ for the majority of well-characterised stars \citep{Bowman2018c*}, which we conclude is observational evidence for IGWs in the photometry of early-type stars.

% % % % % % % % % % % % % % % % % % % % % % % % % % % % % % % % % % % % % % % % 

\section{Discussion and Conclusions}
\label{section: conclusions}

Previous small-sample studies of O and B stars have detected astrophysical red noise in the photometry of these stars (e.g. \citealt{Blomme2011b}). However, the physical origin has, until recently, always been attributed to the assumed presence of a strong, dynamical, clumpy stellar wind. This may be plausible for the most massive stars, but stellar winds is unlikely to explain the stochastic low-frequency variability in late-B and A stars. Furthermore, there is a growing number of upper-main sequence stars for which photospheric variability in the form of IGWs appears to be strongly coupled to the observed wind variability, such that it is becoming increasingly more important to combine photometric and spectroscopic data of massive stars to disentangle the physical mechanisms causing the observed variability (e.g. \citealt{Aerts2017a, Aerts2018a, Simon-Diaz2018a}).

For a sample of O, B, A and F stars observed by CoRoT, the majority of these stars have stochastic low-frequency variability with a parameterisation that is inconsistent with granulation by more than an order of magnitude. Furthermore, the frequencies and the distribution of the measured $\gamma$ values for these stars are in agreement with predictions of IGWs from numerical simulations \citep{Rogers2013b, Rogers2015, Bowman2018c*}. Therefore, from the measured distributions of $\nu_{\rm char}$ and $\gamma$ values shown in Figs~\ref{figure: granulation} and \ref{figure: IGW}, respectively, we interpret the stochastic low-frequency variability in these stars to be caused by IGWs. This is supported by several main-sequence B stars having significant low-frequency variability in their spectra, yet these stars exist within a parameter space on the HR~diagram for which granulation and stellar winds should be negligible. Regardless of the interpretation of the stochastic low-frequency variability found in early-type stars, valuable constraints on the individual amplitudes and frequencies, and the morphology of the background IGW spectrum can be placed for different stars across the HR~diagram. 

In the future, the detection of stochastic low-frequency variability in early-type stars at different evolutionary phases will be possible because of the ongoing TESS mission \citep{Ricker2015}, which was launched on 18 April 2018 and is an all-sky survey mission that will observe hundreds of massive O and B stars. Therefore, the morphology of IGWs across the upper-main sequence can be probed as a function of mass and age for a much larger sample, and provide the observations for the calibration of the next generation of stellar evolution models that include angular momentum transport caused by IGWs.

The work presented here is described in further detail in \citet{Bowman2018c*}.

% % % % % % % % % % % % % % % % % % % % % % % % % % % % % % % % % % % % % % % % 

\section*{Acknowledgments}
The authors thank the organising committees, especially Sylvie and Gerard Vauclair, for such a productive and enjoyable conference. The research leading to these results has received funding from the European Research Council (ERC) under the European Union's Horizon 2020 research and innovation programme (grant agreement N$^{o}$670519: MAMSIE). S. S.-D. acknowledges financial support from the Spanish Ministry of Economy and Competitiveness (MINECO) through grants AYA2015-68012-C2-1 and Severo Ochoa SEV-2015- 0548, and grant ProID2017010115 from the Gobierno de Canarias. Support for this research was provided by STFC grant ST/L005549/1 and NASA grant NNX17AB92G. T. V. R. gratefully acknowledges support from the Australian Research Council, and from the Danish National Research Foundation (Grant DNRF106) through its funding for the Stellar Astrophysics Centre (SAC).

% % % % % % % % % % % % % % % % % % % % % % % % % % % % % % % % % % % % % % % % 

\bibliographystyle{phostproc}
\bibliography{/Users/Dom/Documents/RESEARCH/Bibliography/master_bib.bib}

% % % % % % % % % % % % % % % % % % % % % % % % % % % % % % % % % % % % % % % % 

\end{document}